# Mapping the Artificial Intelligence Divide in Africa: Infrastructure, Accessibility and Capacity

**Abayomi O. Agbeyangi & Jose M. Lukose**
Walter Sisulu University, East London, South Africa
*Corresponding: aagbeyangi@wsu.ac.za*

## Abstract

Artificial Intelligence (AI) has the potential to be transformative for development, but Africa is currently facing a fragmented and challenging “AI divide”. This paper provides an empirical analysis of the current state of the AI landscape and how it compares with Africa's technological preparedness for the future. In our analysis, we approach the “AI Divide” from three angles: infrastructure, accessibility, and human capacity. First, we look at the physical constraints that prevent Africa from integrating digitally. We then evaluate the human-centred factors that limit the development of AI technology on the continent. Finally, we examine the human capacity to develop AI systems on the continent and provide three focused case studies. Our investigation shows that the physical infrastructure needed to build an AI economy on the continent is lagging, with only 38% internet penetration, poor broadband coverage and less than 1% of all data centres globally. Other constraints include high data costs relative to income, gender-based digital divides, and the need to build more representative NLP models that can understand Africa’s native languages. However, there are positive trends towards the emergence of local initiatives and grassroots movements, such as startups and universities, contributing to AI development on the continent. Based on these findings, we provide concrete recommendations to policymakers to help develop a more comprehensive and equitable AI ecosystem on the African continent.

## 1.0 Introduction

Artificial intelligence, along with digital technologies, is capable of transforming the economy by accelerating economic growth, optimising resources, and delivering better public services around the world (Ashok et al., 2022; Feijóo et al., 2020; Oleet, 2025). However, this is not an automatic process that occurs as soon as the technologies appear. They are possible only under certain preconditions related to the developed infrastructure, access to the digital world, and skilled personnel. Unfortunately, all these conditions are less prevalent in Africa than in other developed countries. Thus, there is an "AI divide" that exacerbates the current global inequalities rather than minimises them (de-Lima-Santos et al., 2024).

Even as countries such as South Africa, Nigeria, Kenya, and Egypt are making strides in the area of AI technology, the majority of the African continent is largely marginal to the AI revolution happening at the global level (Malatji, 2026; Walter, 2024). It is more than a question of late development but rather one of basic structural constraints, where the lack of basic infrastructure limits the possibility of sophisticated computing.

The study explicitly puts Africa in the global view of the AI divide. We provide an empirically grounded analysis of the compounding deficits of connectivity, computing power and specialised skills, which severely limit the continent's innovation potential. We measure these gaps using the most recent authoritative evidence, including that from the International Telecommunication Union (ITU) (ITU, 2025a), the World Bank (World Bank Group, 2025), the Global System for Mobile Communications Association (GSMA) (Sajid, 2025), and regional indices (CIPIT, 2025; Oloyede, 2025b), and compare Africa's present trajectory to those of other global regions. We present sub-regional comparisons (comparing West, East, North, and Southern Africa) and differences between Tier-1 tech ecosystems (e.g., Nigeria, South Africa, Kenya) and emerging markets to reflect the continent's diversity.

To systematically investigate this complex landscape, we categorise the African AI divide into three interlinked dimensions:

- Infrastructure: The physical and network hardware needed to support AI. This includes broadband penetration, the geographic distribution of data centres, availability of cloud compute, reliability of the electricity grid and deployment of 3G/4G/5G mobile networks.
- Accessibility: The socio-economic factors that determine how individuals and businesses engage with AI. This includes smartphone and device ownership, the affordability of data, the crucial need for localised Natural Language Processing (NLP) and culturally relevant content, basic digital literacy and the regulatory environments governing data use.
- Capacity: The human and institutional systems necessary to support an AI ecosystem. This includes higher education and specialised AI training, academic research output, the strength of start-up ecosystems, venture capital allocation, and the continued challenge of talent emigration (brain drain).

The rest of the paper addresses these dimensions in a systematic way. Sections 2 to 4 examine infrastructure, accessibility and capacity establishing the quantitative realities of the divide. These macro-level trends are demonstrated by three case studies in Section 5, which show how localised initiatives are navigating these barriers. Finally, Section 6 synthesises these findings to provide actionable and context-specific recommendations in the areas of policy adjustments, infrastructure investments, and strategic partnerships, and proposes priority metrics for accurate monitoring of Africa's progress on the global AI frontier.
.

## 2. Infrastructure Gaps: Connectivity and Compute

### 2.1 Internet and Broadband Access

The fundamental roadblock at the core of the African AI divide is that the continent continues to experience the world's lowest overall internet penetration. While high-speed digital access is treated as a baseline utility in advanced economies, the reality on the ground across Africa is starkly different. In 2024, only approximately 38% of Africans were actively online (ITU, 2024a). This was a slight improvement from about 37% in 2023 (ITU, 2023). This figure is well below the global average of some 68% and even below the averages in many low-income countries outside the region. In particular, the estimated Internet use rate in Africa in 2025 was 36%, compared to 74% globally, and well below regions such as Europe, CIS and the Americas, where Internet use rates ranged from 88 to 93% (ITU, 2025c). However, this macro-statistic masks profound geographic and socioeconomic inequities. The landscape is characterised by a deep rural-urban divide; urban internet usage far exceeds rural usage by a huge margin (estimated at 83% to 48%) (Mediacentre, 2024) pointing to severe infrastructure coverage gaps outside of metropolitan areas. As fixed-line broadband is almost non-existent for the average household, mobile networks are the main gateway to the digital world. But even mobile internet is still a luxury for most. GSMA (2024) reports that 60% of the population in Sub-Saharan Africa does not use mobile internet, despite living within mobile broadband network range, which it refers to as a "usage gap" and cites financial, literacy and device-related barriers as the reasons for this.

Taking a closer look at specific country-level data, the disparities become even more pronounced. The continent is essentially split between mature telecoms markets and weak, infrastructure-deficient regions. As shown in Table 1 and Figure 1, countries with large youth populations such as Nigeria and Kenya, often touted as vibrant startup ecosystems, still have less than half of their population online. Unlike Morocco (92.2%) and South Africa (79.6%), which have benefited from decades of consistent institutional and private investment in telecommunications infrastructure. Regionally, Southern and Northern Africa have the highest connection rates at 64% and 68%, respectively, as of 2022 (CRE, 2022). On the other hand, Central and East Africa are still very limited, with a penetration rate of around 25%. Besides these connectivity metrics, "deficit of network-supporting infrastructure", including terrestrial networks such as fibre, particularly in rural areas, is mentioned as a reason for low Internet penetration (Falowo & Falowo, 2026). The absence of high-speed backbone connectivity makes it difficult to implement bandwidth-intensive technologies, as mobile devices are often the sole access point, despite the high costs of data and modern hardware (Ehimuan et al., 2024).

**Table 1:** Internet use and data centre presence in selected African countries.

| Country/Region | Internet Users (% of population) | Data Centres (#) | Extent |
|---|---|---|---|
| Africa (Total) | 38% (2024) (ITU, 2024a) | >140 (continent-wide) (CIPIT, 2025) | Houses <1% of global data centres (Filipenco, 2026) |
| Nigeria (West) | 45.5% (October 2025) (Kemp, 2025c) | 28 (Datacenters, 2026a) | 227M population, ~69.2% mobile penetration (Kemp, 2025c) |
| Kenya (East) | 48.0% (Jan 2025) (Collins, 2025) | 19 (Datacenters, 2026a) | ~70% smartphone penetration (Collins, 2025) |
| South Africa (South) | 79.6% (October 2025) (Kemp, 2025d) | 64 (Datacenters, 2026a) | 3 largest cloud regions (AWS/Azure/GCP); 4G/5G widely deployed |
| Egypt (North) | 82.7% (October 2025) (Kemp, 2025a) | 13 (Datacenters, 2026c) | Major internet hub; heavy undersea cable connectivity |
| Morocco (North) | 92.2% (October 2025) (Kemp, 2025b) | 14 (Datacenters, 2026a) | Highest penetration; advanced ICT development |

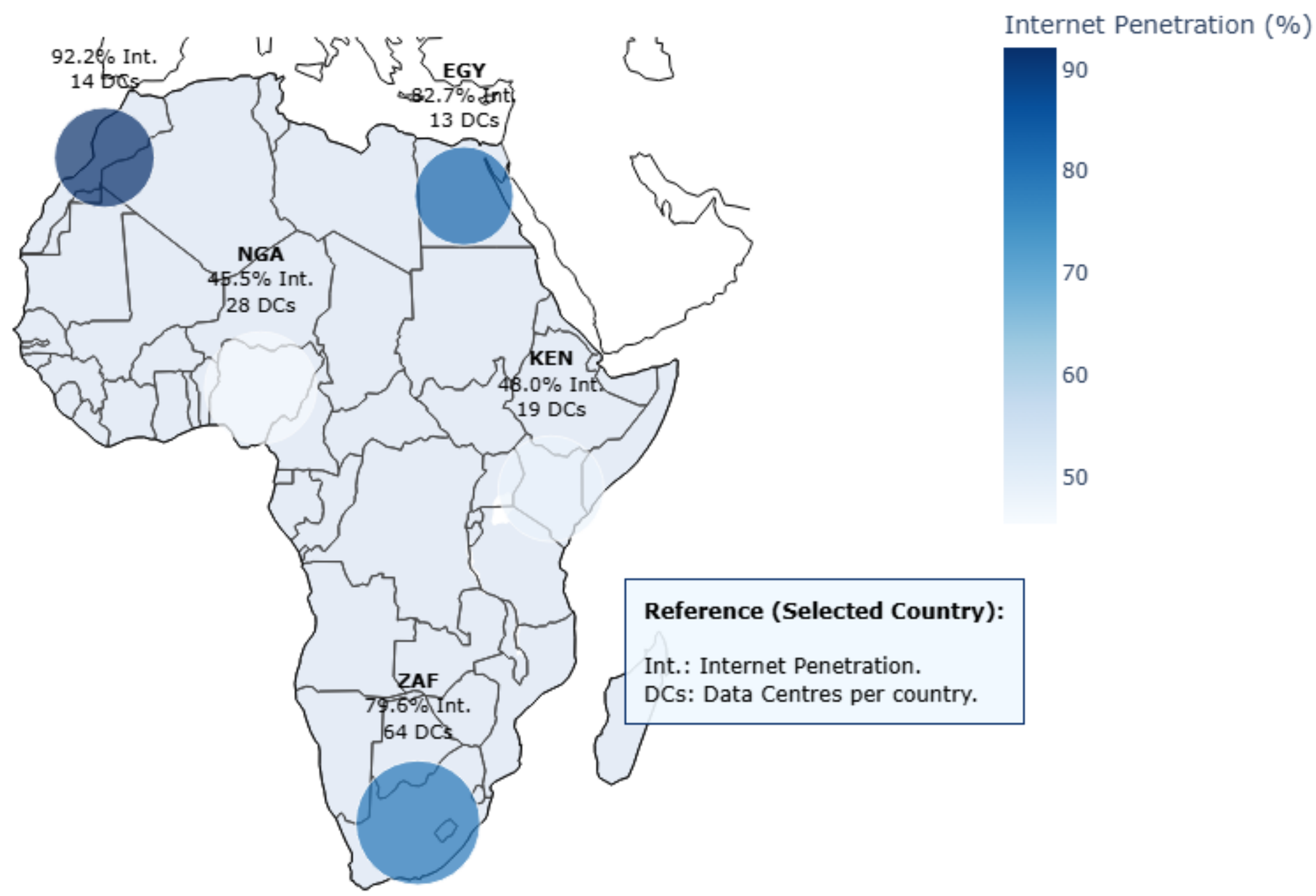

**Figure 1: Internet Penetration and Data Centre Concentration.** A geographic comparison of foundational digital infrastructure across key African markets. Bubble size represents the absolute number of data centres per country, while the colour gradient reflects the percentage of the population with active internet access.

**Source:** Author.

These connectivity deficits are tied directly to issues of quality and affordability. According to World Bank assessments as of 2022, while the availability of formal broadband (≥10 Mbps) crawled upward from 26% in 2019 to 36% by 2022, actual connection speeds remain frustratingly low, with a median broadband speed across Africa of just 8.2 Mbps (World Bank Group, 2023). Furthermore, data remains financially punitive. According to Mwema and Birhane (2024), in 2023, only 37% of the African population used the internet, generating only 130 GB of traffic per month, and the African digital divide was lower than the global average across all aspects, "especially with significantly low broadband connectivity". Mobile broadband prices average roughly 5% of the monthly Gross National Income (GNI) per capita for a meagre 1GB data allowance. This is more than double the UN Broadband Commission's target[1] threshold of 2% or less, making routine internet usage economically unfeasible for the working poor. Beyond these fiscal constraints, the lack of transparency in undersea cable governance and ownership structures has further hampered the expansion of reliable, high-capacity bandwidth across the continent (Mwema & Birhane, 2024). Additionally, the reliance on mobile technology to bridge these gaps has proven insufficient to support the high-speed requirements of modern digital economies, leaving many regions unable to move beyond basic connectivity.

Finally, the physical deployment of high-speed networks is still very fragmented. In 2024, the 4G network coverage was only 52% of the population in Africa (ITU, 2024a), while the 5G infrastructure coverage was 11% (ITU, 2024b). This is in contrast to the 5G coverage in high-income countries, where it is almost universal. As of 2025, the deployment disparities are still evident, as the high cost of network expansion in rural, sparsely populated areas continues to discourage private capital investment (Xulu & Malatji, 2025). In deep rural areas, the coverage gap becomes total exclusion: some 14% of the continent's rural population has no mobile broadband access (ITU, 2024b). While massive undersea fibre-optic consortia and terrestrial network expansions (e.g. Meta's 2Africa and Google's Equiano cables) (Agwor et al., 2025; CRE, 2022), are actively landing on African shores, they are encountering severe "last-mile" distribution bottlenecks. High domestic backhaul costs and fragmented local grids mean that the huge bandwidth arriving at coastal ports rarely reaches the inland communities and local institutions that most desperately need AI innovation.

[1] https://www.broadbandcommission.org/advocacy-targets/3-universal-broadband/

### 2.2 Data Centres and Cloud Infrastructure

While the initial challenge of basic internet access still exists, the real bottleneck for artificial intelligence is computing capacity. The development of AI, in particular the training and deployment of sophisticated machine learning models, requires massive, localised data processing capabilities. Here, Africa's computing infrastructure is critically scarce. As discussed in Section 2.1, the continent currently accounts for less than 1% of the total data centre capacity in the world (Filipenco, 2026). Industry experts estimate that Africa needs hundreds of new facilities to meet the rising demand for digital services and to build a self-sustaining AI ecosystem. Recent industry listings suggest that there are only about 256 formal data centres in operation across the entire continent (Datacenters, 2026a), which is far less than the capacity of a single major Western nation, like the United Kingdom, with 540 datacentres or Germany, with 520 (Datacenters, 2026b).

Furthermore, this limited infrastructure is highly concentrated in a few key markets:

- **South Africa:** Dominates the continent with 64 established centres.
- **Nigeria:** Follows as a secondary hub with 28 centres.
- **Kenya:** Serves as the East African hub with 19 centres.
- **Morocco & Egypt:** Anchor the North African market with 14 and 13 centres, respectively.

These numbers may not reflect smaller, informal or private enterprise facilities, but the fact is that the majority of African nations do not have a single modern, Tier-III data centre. African data centre development has made great progress, with over 140 colocation and hyperscale data centres now in operation, but the majority of these are in Nigeria, South Africa, Egypt and Kenya (CIPIT, 2025). For example, Main One, a connectivity and data centre provider, has expanded its facilities to include a Tier-3 data centre in Appolonia, Accra (Ghana), and the ABJ-1 data centre in Abidjan (Côte d'Ivoire), which has increased regional connectivity. Historically, Africa's critical multi-country shortage of localised compute power has indicated that African enterprises and researchers have had to rely on offshore cloud regions (usually Europe or the United States). This dependency causes severe latency problems, increased costs for data in transit and complex data sovereignty issues. Until recently, global hyperscalers like Amazon Web Services (AWS) and Microsoft Azure had no physical presence in Africa apart from South Africa (CIPIT, 2025).

Importantly, this landscape is beginning to change. Microsoft opened its first African cloud region in Cape Town in 2019, and then extended Azure into Lagos, Nigeria. AWS has likewise set up shop in Cape Town, and Google Cloud is aggressively expanding its footprint. However, despite these high-profile investments, access to specialised, high-performance computing is still very limited outside of Tier-1 tech hubs, especially the Graphics Processing Units (GPUs) and specialised AI servers required for advanced model training (CIPIT, 2025; Filipenco, 2026).

Finally, the expansion of physical compute infrastructure is inextricably linked to the continent's energy grid. Data centres consume vast amounts of energy, and reliable electricity is a major bottleneck in the development of AI. According to the GSMA, African tech companies consistently rate volatile power cuts and voltage fluctuations among their chief operational concerns (GSMA, 2024). The energy landscape is highly fragmented: although North Africa, South Africa and Kenya have relatively stable grids, blackouts are a daily reality in much of West and Central Africa. Even South Africa has a history of scheduled grid blackouts (load shedding). In the end, that difference in dependable baseload power often determines where global tech giants are willing to build, further deepening the AI divide.

### 2.3 5G Deployment and Edge Connectivity

While high-income countries are fast transitioning to 5G as the baseline for next-generation digital economies (with about 84% population coverage (ITU, 2025b)), the mobile connectivity landscape in Africa is structurally different. Currently, only 10%–15% of the continent's population is covered by a 5G service footprint (ITU, 2025b). According to the report (ITU, 2025b), 74 per cent of the population in Europe is covered by 5G, placing the region at the forefront, followed by the Asia-Pacific region with 70 per cent and the Americas region with 60%. In contrast, coverage is considerably lower in the Arab States region, at 13%, Africa at 12% and the CIS region at 8% of the population covered by 5G. Furthermore, this small number is heavily biased towards wealthy, urban centres, mostly in South Africa, Morocco, Egypt, Kenya and Nigeria.

For most of Sub-Saharan Africa, however, 4G rather than 5G remains the frontier technology. In 2024, the 4G coverage areas covered only around 65% of the population, with almost half of the continent being either on the legacy 3G or completely disconnected (Shanahan, 2025). Such dependence on legacy network architectures creates a severe bottleneck for the deployment of edge computing. For AI applications that require real-time, low-latency processing (such as autonomous logistics or advanced agricultural monitoring), edge computing, placing small-scale data centres closer to end users, is essential. While these edge-hosted AI services are cautiously being piloted in mature markets such as Kenya, Nigeria and South Africa, the wider continental ecosystem remains in its infancy. In response to terrestrial network failures, especially in deep rural areas, there is an increasing reliance on low-Earth orbit (LEO) satellite providers (for example, Starlink) and microwave links (Shanahan, 2025). But while these alternatives are novel, they often lack the cost-effectiveness required to support high-density AI workloads at consumer scale.

All of these factors combined mean that Africa's basic AI readiness lags behind Asia, Europe and the Americas. Africa, in particular, suffers from low broadband penetration, a clear lack of localised data centres and intermittent baseload power. It is important to realise that this landscape is not static; there is a tangible surge of capital injection. Undersea cables, such as Google's Equiano and Meta's 2Africa, are vastly expanding coastal bandwidth through mega-projects, while regional coalitions such as the Africa Data Centres Association are successfully facilitating domestic compute investments (CIPIT, 2025; Filipenco, 2026). But the scope of the challenge left is enormous. Multilateral development trackers estimate that bridging this foundational divide will require an additional $100 billion in broadband infrastructure investment alone over the coming decade (Acquaye, 2025; CRE, 2022). Until

these physical constraints are addressed, widespread equitable deployment of artificial intelligence across the continent cannot be fully realised.

## 3. Accessibility Barriers

### 3.1 Device Ownership and the Affordability Crisis

While infrastructure dictates whether a network even exists, accessibility defines who can access it. The most significant issue in Africa seems to be the lack of compatible devices (Agbeyangi & Suleman, 2024). As mentioned by ITU, the ownership of mobiles in poor countries is estimated at approximately 53%, drastically differing from a stunning 97% found in developed countries (ITU, 2025d). However, to ensure effective AI adoption, the important parameter remains smartphone prevalence.

To use AI applications, ranging from advisory services to basic generative tools, one needs to have a smartphone (Mannuru et al., 2025; World Bank Group, 2025). While there are estimations of finally reaching the 50% prevalence level in the region's smartphones (CIPIT, 2025), the vast majority of devices in rural areas remain those featuring a few basic functions. Thus, what emerges as an ironic statistics problem is the fact that, despite having mobile connection rates exceeding the number of people in some African countries (multi-SIM), internet prevalence appears to be much lower (Falowo & Falowo, 2026; Nye & Nye, 2014). For instance, in Kenya, although the prevalence rate exceeds 100 phones per 100 citizens, only 48% of residents are connected to the internet (Collins, 2025).

This gap in usage is mainly due to the cost of connectivity (Robinson et al., 2020). In a number of regional markets, the purchase of an entry-level smartphone is an unaffordable percentage of the average citizen’s monthly income. Secondly, the costs of recurrent data are exorbitant. The World Bank notes that the cost of purchasing just 1GB of data in 2023 is about 5% of the monthly Gross National Income (GNI) per capita across the continent (World Bank Group, 2023), which is more than twice the UN Broadband Commission’s affordability target of 2%. Even something as simple as an internet search or a messaging app can cost the poorest demographic a fortune in household spending. Consequently, the computational resources required to operate advanced AI systems, which often require high-speed and stable data connections, remain inaccessible to a large segment of the population (Asiedu et al., 2025).

Regulatory approaches to this affordability crisis are highly fragmented. While in recent years, proactive nations like South Africa[2] and Mauritania[3] have structurally reduced mobile taxes to stimulate adoption, other governments continue to levy aggressive VAT and import duties on digital devices, further stalling the democratisation of AI. Moreover, this fiscal environment is exacerbated by inconsistent regional ICT policies, which result in tariff-laden infrastructure imports artificially inflating service prices (Selemela et al., 2023; Velasco et al., 2026).

### 3.2 Language and Localisation Deficits

---

[2] https://www.connectingafrica.com/regulation/south-africa-plans-to-stop-import-tax-on-low-cost-smartphones

[3] https://www.ecofinagency.com/news-digital/1103-53675-mauritania-cuts-phone-taxes-but-new-controls-raise-access-concerns

The global digital economy is largely founded on a small number of dominant languages (English, French, and Chinese). In contrast, over 2,000 languages are spoken in Africa (Wawire et al., 2025). This mismatch in language is a serious bottleneck to accessibility: the vast majority of AI interfaces, training datasets and outputs are completely inaccessible in widely spoken languages like Hausa, Amharic or Shona. This leads to poor performance of existing Large Language Models on non-English regional dialects, causing a substantial "data poverty" that limits the relevance and accuracy of AI applications for indigenous communities (Segun et al., 2025). The lack of high-quality and representative datasets for these low-resource languages also limits the development of culturally nuanced machine learning models by developers (Folorunso et al., 2024).

If localisation does not happen, this will mean that large sections of the population, especially older, rural, and less educated individuals, are automatically excluded. Although the global hyperscalers such as Google and Microsoft are now starting to develop AI applications in major African languages, the levels of availability of these localised applications are extremely limited. Moreover, there is also a paradox regarding the pricing strategies, which are uniform for all developing countries but significantly higher than local purchasing power in terms of the cost of accessing advanced technologies.

Notably, the most successful attempts at solving this problem come from Africa itself. Various developer communities are actively engaged in creating datasets that are often neglected by the leading Western technology companies. One such attempt comes from the Masakhane NLP project[4], which managed to create more than 400 AI models and 20 datasets tailored specifically for African languages (CIPIT, 2025). Such indigenous initiatives will prove to be crucial in making sure that future AI tools can speak to Africa's people.

### 3.3 Digital Literacy and the Gender Divide

Hardware and localised software are impractical without the basic digital literacy to use them. Many African communities still have a shockingly low level of confidence in their ability to navigate devices, applications and the internet. UNESCO and the World Bank report that most rural Africans have not undergone formal ICT training (Gala et al., 2025; UNESCO, 2025). Educational systems are slowly modernising, with African nations actively integrating computer labs into primary education and universities through expanding data science curricula, but the informal sector lags behind. Small business owners and smallholder farmers, surveys consistently show, lack the basic digital fluency to take up AI-driven enterprise tools or automated agricultural advisory services (O'Neill et al., 2024).

Furthermore, this literacy gap is very gender-specific. In the continent, women are far less likely to own a smartphone or use the internet (Mulungu, 2025; World Bank Group, 2023). The digital gender gap is rooted in systemic educational inequities, gaps in income independence, and restrictive social norms. While targeted interventions, such as UNESCO's "Girls in ICT" programmes, are working to redress these inequities, progress remains uneven, threatening to leave African women disproportionately excluded from the benefits of the AI economy (UNESCO, 2026).

### 3.4 The Regulatory and Policy Environment

---

[4] https://www.masakhane.io/masakhane-african-languages-hub

The availability of AI does not only rely on economic and educational considerations but also on the digital policies that exist in this regard. Presently, there is a clear policy awakening being witnessed on the continent. In total, about 39 out of the 54 countries in Africa have enacted legislation that deals with data protection (Oloyede, 2025). In addition to that, a group of pioneering countries, such as Egypt, Ghana, Kenya, Benin, and Algeria, are working on developing national AI strategies for their respective countries (OECD, 2025). Generally, these policies tend to be quite ambitious and tailored towards achieving specific goals that are aligned with social and economic growth. The case of Ghana is quite interesting in that the country's National AI Strategy 2023-2033 focuses on using AI to achieve financial inclusivity via fraud detection technology.

Nevertheless, an important policy gap persists on the continent. There are many countries that rely on obsolete digital policies or lack any coherent AI governance altogether. This regulatory ambiguity in terms of international data transfers, surveillance programs, and copyright enforcement through algorithms has a very real effect and dissuades both local innovations and international investments in AI technologies. Fortunately, there are efforts made towards achieving harmonisation in policy. In 2024, the African Union released an all-encompassing strategy for the use of AI on the continent, calling for member states to strengthen their institutional capacity in addition to developing ethical guidelines for AI usage (OECD, 2025; Unver, 2025). This strategy focuses on the development of national AI strategies within member states, building regulatory frameworks, and implementing ethical guidelines for AI. Provided that these continental guidelines are properly domesticated by member states, they can go a long way in harmonising the regulatory environment.

In particular, the idea that the issue of the AI divide should be approached only from the perspective of the underlying physical infrastructure is an equally important misunderstanding. Even in cases where there is enough capacity in terms of broadband internet and computing clusters, regular Africans would have to contend with many more human-made accessibility hurdles. This might include, among others, the shortage of inexpensive smartphones, excessively high costs for repeated access to data, the absence of language-relevant AI algorithms, deep-rooted illiteracy, and a divided policy landscape. This list of potential accessibility impediments cannot be separated from the hardware component of the digital divide, as both feed off each other. Addressing the issue effectively requires a broad approach, ranging from governmental efforts aimed at subsidising devices and lowering data tax to funding of local language training databases and digital literacy education programs. As illustrated in Figure 2, neglecting to tackle accessibility would mean that investing in physical infrastructure and submarine cables is nothing but increasing inequality through digitisation.

## 4. Capacity Gaps: People and Ecosystems

### 4.1 Education, Skills, and the Brain Drain

Human capital is the key element that makes up the AI-driven economy, but, unfortunately, Africa is yet to develop a robust formal AI education infrastructure (Ahmed et al., 2025). While there are a few standout institutions, such as the African Institute for Mathematical Sciences (AIMS)[5], offering specialised data science master's degrees, and universities like Africa University[6], the University of

[5] https://aims-network.org/

[6] https://africau.edu/program/bachelor-of-science-honours-in-artificial-intelligence/

Cape Town (UCT)[7], and the University of Johannesburg (UJ)[8], which run dedicated AI undergraduate programmes, the output of graduates is microscopic relative to the continent's demand. Conversely, Nigeria and South Africa have a massive and rapidly growing AI ecosystem, with many top-tier universities offering postgraduate and specialised research programmes in Artificial Intelligence, but they need to strengthen this at the undergraduate level.

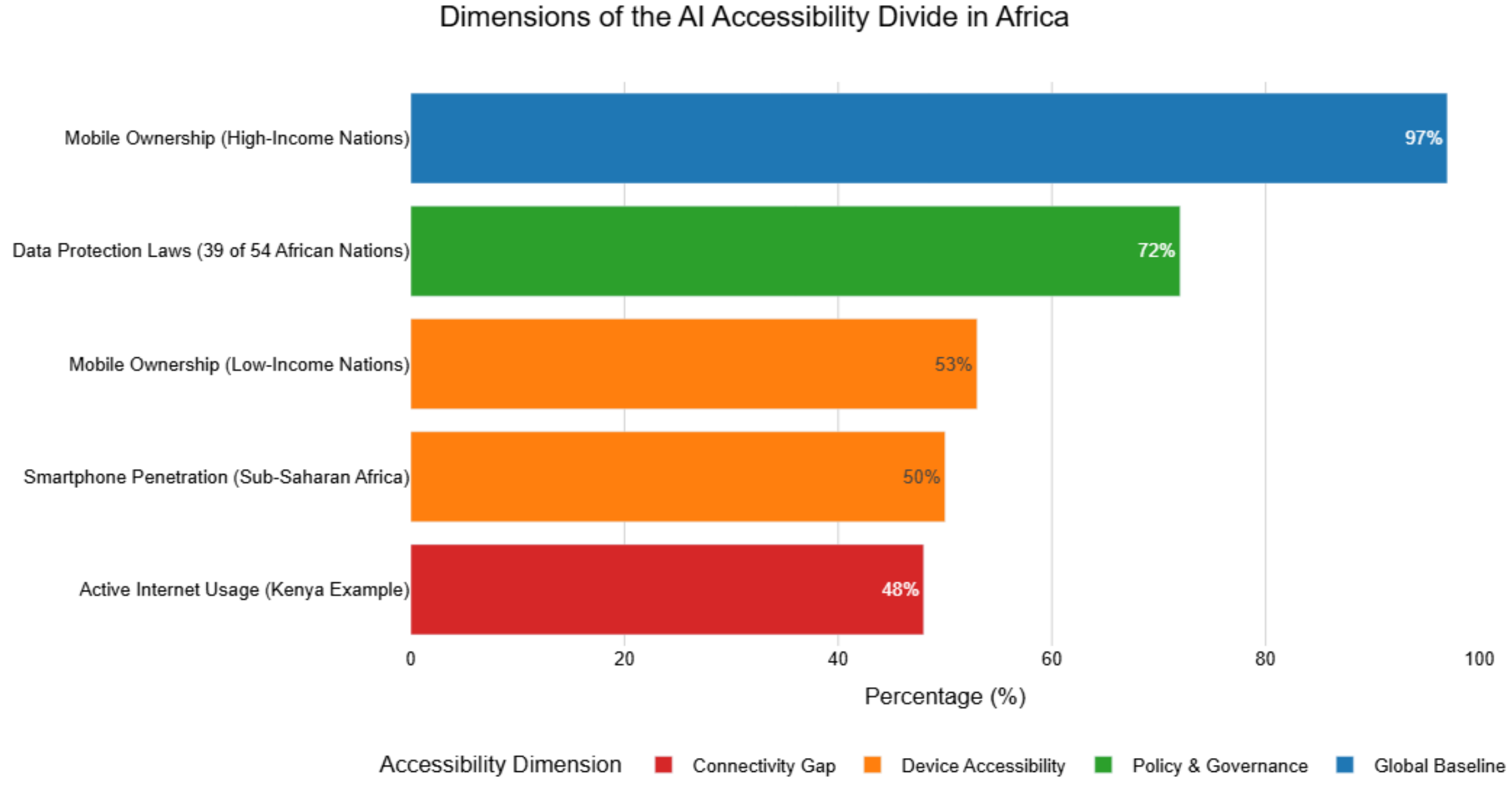

**Figure 2. Dimensions of the AI Accessibility Divide in Africa.** The figure highlights disparities in device ownership, internet connectivity, and policy readiness that influence access to AI technologies across Africa.

**Source:** Compiled by the authors from (ITU, 2025b), (CIPIT, 2025), (Collins, 2025), and (Oloyede, 2025).

This scarcity of talent is actively hindering economic growth. Recent corporate surveys paint a critical picture: SAP reports that 90% of African organisations say a lack of AI expertise is already negatively impacting their business operations (Madikane, 2026; Stewart, 2025). Recognising this deficit, another SAP study revealed that two-thirds of regional firms are now prioritising emergency workforce reskilling and upskilling in AI by 2025 (Lechman, 2026).

While the continent's overall Science, Technology, Engineering, and Mathematics (STEM) enrolment is steadily growing (UNESCO still feels Africa needs an additional 23 million STEM graduates by 2030 (UNESCO, 2024)), very few specialise in advanced machine learning. Instead, the bulk of tech talent naturally gravitates toward more established, immediately lucrative fields such as software development or telecommunications.

Furthermore, the ecosystem suffers from a profound "brain drain” (Kanyesigye, 2026; Walter, 2024). African ICT specialists, data scientists, and academics frequently emigrate to Europe or North America in search of advanced research facilities, stable funding, and higher compensation.

[7] https://ai.uct.ac.za/articles/2026-03-26-uct-lead-africas-first-higher-education-dedicated-ai-compute-initiative

[8] https://www.uj.ac.za/university-courses/bsc-in-computer-science-and-informatics-specialising-in-ai-artificial-intelligence/

Consequently, there are relatively few professors focused on AI remaining on the continent. For example, a shortage of tech-ready domestic professionals, including science teachers, data centre engineers, network architects, and cloud operations specialists, is a critical constraint on effectively utilising the region's digital infrastructure (CIPIT, 2025; Manouan et al., 2026). While recent "brain gain" initiatives and diaspora fellowships attempt to reverse this trend, talent attrition remains a significant, systemic leak in the African capacity pipeline.

### 4.2 Research and Innovation

Reflecting the talent scarcity, Africa's formal research output in AI remains modest. Africa contributes less than 5% to the global pool of AI research outputs, indicating a notable underrepresentation in top-tier AI advancement (Khan et al., 2024; Manouan et al., 2026). Specifically, there is a critical shortage of tech-ready domestic professionals, including academic professionals, hindering the effective utilisation of existing digital infrastructure for AI research output. Furthermore, less than 25% of African higher education students pursue STEM qualifications, further limiting the pipeline of future tech professionals (Manouan et al., 2026).

This research is not evenly distributed; it is highly concentrated within a few academic strongholds. North Africa, led by countries like Tunisia, Egypt, Algeria, and Morocco, and Southern Africa, led by South Africa, West Africa, with Ghana and Nigeria showing potential and Kenya, accounts for the vast majority of the continent's AI papers (CIPIT, 2025). Furthermore, collaboration networks tend to bypass intra-African partnerships; researchers are much more likely to co-author papers with European or Asian institutions (where funding originates) than with peers in neighbouring African nations (Aryee et al., 2026; Pouris & Thirion, 2024).

Despite the lack of institutional funding, grassroots innovation is thriving. Annual conferences like the Deep Learning Indaba[9] and vibrant peer communities like Masakhane[10] have successfully galvanised a new generation of young African AI researchers. This is because these communities prioritise the practical application of artificial intelligence, creating solutions specific to their current environment, including optimising farming, financial services, and healthcare diagnostics. However, without access to competitive, large-scale grants, these brilliant proof-of-concept projects rarely scale into enterprise-grade applications.

### 4.3 Funding and Entrepreneurship

The financial framework that would make AI research a profitable business in Africa is considered to be relatively weak. Firstly, the cost of building, training, deploying, and maintaining sophisticated AI systems, such as cloud computing costs, represents an enormous financial burden, especially to smaller companies and startups (OECD, 2025). Secondly, the flow of capital in the AI landscape in Africa has mostly relied on foreign direct investment from major technology companies in order to develop infrastructure (CIPIT, 2025). Although the investment in African AI start-ups in the form of venture capital has indeed increased significantly since 2019, with over $1.25 billion reportedly being invested during that time period (Oloyede, 2025), it represents only a negligible fraction of the annual worldwide AI investments, which are known to consistently exceed $100 billion per year (Qaleon, 2026). The US is at the top of the list, making up about 56% (around $124 billion) of the total number of venture capital investments globally (OECD, 2026).

---

[9] https://deeplearningindaba.com/

[10] https://www.masakhane.io/

Furthermore, in the African region, the capital is extremely centralised. In line with the findings of Okolo (2026), investments that go into African AI startups usually comprise an insignificant part of the total sum of funds globally. In particular, the "Tier-1" hubs – Nigeria, Kenya, South Africa, and Egypt – have access to virtually all funding. Meanwhile, "Tier-2" countries such as Morocco and Ghana got around $160 million in VC investments, with Tier-3 states receiving almost nothing (Oloyede, 2025).

The source of this investment is also problematic from the point of sustainability. The ecosystem is overly dependent on external sources of technology finance, international grants by foreign non-profits, and international agencies, including such organisations as the World Bank, AfDB, and UNDP. At the same time, African venture capital is evolving rapidly and outgrowing its stage of “infancy” (Kabengele & Hahn, 2025). According to Mkalama & Ouma (2025), despite the increase in the total sum of VC investments made in Africa, the continent is still far from being an equal player in terms of attracting venture investments; its share does not exceed 2%. In addition, apart from some cases, for example, the AI “grand challenge” funded by South Africa’s government, African countries do not allocate significant sums from their national budgets to AI innovations (OECD, 2025).

### 4.4 Startups and Ecosystems

Despite all these challenges, there are still resilient signs of a vibrant startup environment in Africa. According to industry watchers, there are hundreds of AI-focused technology startups currently operating in Africa (Oloyede, 2025a; Wereko-Brobby, 2025). As per the mapping done in 2025, most of them have been found in South Africa (~63), Kenya (~59), Nigeria (~57), and Egypt (~16) (Oloyede, 2025a). The problem is that the term "AI startups" has been found to be somewhat complex when referring to the startups in Africa. These startups do not develop any foundational AI models (as compared to companies like OpenAI or Anthropic), but they use AI technology in different sectors. They are basically "AI-enabled" enterprises which integrate ML APIs on their existing applications. Examples include Kenya’s PBR Life Sciences standardising regional health data, South Africa’s Aerobotics deploying AI for precision agriculture (CIPIT, 2025; Wereko-Brobby, 2025).

Scaling such challenges may prove difficult. In addition to the problem of seed financing, there are other problems, such as slow bureaucracy within the business environment, slow company integration processes, and ineffective intellectual property rights protection (Kato, 2026). Although technology accelerators (such as CcHUB[11] or Nairobi Garage[12]) offer much-needed assistance, they remain concentrated mainly in the capitals of the respective countries. Moreover, the issue of “spin-offs” from local universities is conspicuous by its absence due to the inflexibility of university environments (Manouan et al., 2026).

Generally, there is a great deal of energy when it comes to innovation in Africa through artificial intelligence, yet the structure is still embryonic in nature. Comparing the highly dynamic cities of Lagos, Cape Town, and Nairobi, which have thriving AI communities, with countries like Chad and the DRC, where AI development is virtually non-existent, demonstrates the significance of sub-regional issues that impact capacity, including the availability of educational programs, financial resources, and appropriate policies.

[11] https://cchub.africa/
[12] https://nairobigarage.com/

## 5. Case Studies

The following case examples provide practical illustrations of how some of these broad concerns manifest themselves at the micro level by showcasing three distinct projects that have been instrumental in helping Africa establish its AI ecosystem.

### Case 1: Government Initiative – South Africa's AI Institute and National Policy Push

As the continent's leader in digital economy, with around 70% of the total digital economy capacity of Africa, South Africa has opted for a very aggressive state-driven way of bridging the AI capacity deficit by forming the Artificial Intelligence Institute of South Africa (AIISA) and developing the regulatory environment within the nation (Department of Communications and Digital Technologies, 2026; Mpedi et al., 2022). The nation did not centre its actions on any particular city but collaborated with different educational institutions to form several hubs based on the sectors in which they operate (OECD.AI, 2025). Such institutions include the Central University of Technology (CUT) AI hub, which mainly deals with agricultural and tourism industries, and the Defence Artificial Intelligence Research Unit (DAIRU), founded in 2024 at the Military Academy in Saldanha (Jones, 2026).

The development of this infrastructure is increasingly being supported by good governance. The South African Cabinet adopted the Draft National AI Policy in 2026, which suggests that there is now an intention to introduce accountability mechanisms into the AI landscape through supervisory structures, as opposed to setting up an AI regulatory authority (Department of Communications and Digital Technologies, 2026). This policy is conscious of the challenges posed by a growing digital divide, which has led to considerable investment in computing power for local innovators.

The South African government's effort is notable because it does not concentrate merely on technology. Indeed, the policy framework for 2026 also includes efforts at preserving culture and human development. In terms of educational development, there are plans included in the framework for implementing AI technology across all levels of schooling in South Africa. Moreover, the framework calls for the implementation of AI technology in the digitisation of marginalised knowledge systems, especially the 12 indigenous languages of the country.

### Case 2: Commercial Startup – PBR Life Sciences

Established in 2016, PBR Life Sciences employs innovative practices by leveraging artificial intelligence to process, collate, and standardise healthcare data from Africa (Praxis, 2026; Wereko-Brobby, 2025). The firm operates mainly between Lagos and London and discovered that although pharmaceutical firms globally were interested in operating within new markets, they lacked access to reliable clinical information in Africa (PBR Life Sciences, 2026).

PBR developed proprietary AI models able to unify chaotic and unstandardized clinical and pharmacy data into standardised WHO data forms. The innovation had a dramatic effect: it made the process of processing standard data take 20 minutes instead of the previous eight-to-nine-month period (Wereko-Brobby, 2025). As part of recent efforts to expand into Ghana and other nearby markets, PBR joined the Startup Accelerator for Health by Google (Omoniyi, 2024). Moreover, the firm plans to introduce a "Health Data Lab" to offer anonymised data sets representing black communities for further drug research on a global level (Ashiru, 2025). Thanks to a $1 million investment in pre-seed funding from reputable sources, including Techstars and Launch Africa, PBR has proven itself a fast-growing enterprise. On the other hand, the very successful path chosen by PBR demonstrates the immense possibilities of indigenous AI technology as well as its inherent limitations in the context of an emerging market like Africa (Mkalama & Ouma, 2025).

### Case 3: Research and University Ecosystem – Masakhane NLP Initiative

Masakhane[13] is a unique Pan-African and open-source initiative formed in 2018, whose main objective is to create NLP technologies dedicated to African languages (Cashman, 2020). In a few years since its inception, this initiative managed to deliver outstanding results in terms of creating more than 400 language models and more than 20 datasets, including Luganda, Yoruba, isiZulu, and Amharic[14].

This project, working mostly outside the bounds of existing, highly funded institutions, addresses the issue of localisation head-on. Masakhane has already trained more than 100 African data scientists and continues to foster growth by participating in the university community through events such as the Deep Learning Indaba conference. The project has continued to grow in its efforts, with the Masakhane African Languages Hub presently collecting high-quality multimodal data for 50 languages in an attempt to collect 500 hours of voice data per language for voice-to-voice translation (Chair, 2026).

Realising the importance of this kind of community engagement, several big organisations, such as the Gates Foundation, Google.org, and the Microsoft AI for Good Lab, have now come together with Masakhane for a project entitled LINGUA Africa that offers considerable financial and cloud computing assistance for initiatives related to language AI (Slator, 2026). The continued success of Masakhane demonstrates one key takeaway point: networked communities can overcome challenges of resources to establish important capabilities, despite a lack of official state-level investment in AI technologies.

## 6. Implications and Recommendations

### 6.1 Implications of the AI Divide

The core deficiencies highlighted in this paper have serious repercussions on the potential future of the digital economy in Africa. In the case of failing to address these interdependent deficiencies, the AI divide may cause further exponential inequality, making the continent forever dependent on advanced countries. There is a likelihood of African countries being consumers of AI solutions that do not align well with their cultural values, rather than developing their own innovations, which could seriously hinder any form of digital independence in the region. On the other hand, through strategic intervention by the stakeholders, there will be a great possibility of surpassing these old systems, as an indigenous AI landscape can revolutionise the socio-economic status of the continent.

In addition to macro-economic sovereignty, the cost of increasing inequality in AI goes deep into the lives of people. The fact is that if we do not take deliberate action to create an inclusive economy based on artificial intelligence, it will only exacerbate the problems that already exist today, especially those related to rural communities and women's groups that are structurally marginalised. Critical life-supporting applications, from early diagnosis to farm automation services, will stay locked away behind the wall of affordability, digital literacy, and language barriers. In this scenario, thousands of individuals will have no opportunity to become part of the new economy and find themselves living on the edge of a world system that fails to understand their situation and serve their interests. On the contrary, putting human-centred capacity development first will make AI a truly democratizing technology for people everywhere.

---

[13] https://www.masakhane.io/

[14] https://www.masakhane.io/ongoing-projects/masakhane-mt-decolonise-science

### 6.2 Priority Recommendations

To bridge this divide and capitalise on the continent's potential, policymakers, investors, and academic institutions must take coordinated action across several critical areas.

**Expanding and democratising connectivity**: Both the government and the private sector should urgently focus on making broadband expansion a priority within underdeveloped and rural regions. This should be achieved through subsidising the cost of erecting towers in rural areas by the government, Wi-Fi in municipalities, and planning to extend the reach of sub-sea cable and fibre infrastructure in the inland region. As part of ensuring cheaper connectivity for consumers, the governments should lower import tariffs and taxes charged on mobile data and phones. In the meantime, efforts towards coordination, such as East African broadband projects (O'grady Vaughan, 2024), and adoption of new technologies, such as satellites or TV white spaces, will go a long way towards bridging gaps in connectivity. Progress in this area can be monitored by tracking internet users as a percentage of the population and assessing regional mobile broadband coverage.

**Investing in data centre and cloud capacity**: The stakeholders need to harness public-private partnerships to hasten the development of the local cloud infrastructure. The governments can entice big players into their countries through the creation of data centre parks at the national level that will incorporate various tax benefits, just like Nigeria's growing carrier-neutral parks[15] or even Konza Technopolis[16] of Kenya. The local banks, including other institutions such as the pension funds, should be aggressively encouraged to finance the critical technological facilities. Additionally, the African countries will need to partner with each other regionally (for instance, West African or Southern African cloud) in order to share the huge investment cost necessary for building the computer infrastructure.

**Supporting affordable devices and services:** Policymakers should look to the implementation of proven models like Ghana's policy on giving away free laptops to school teachers[17] or Ethiopia's deeply discounted data packages[18]. The creation of a well-regulated market for used smartphones and the development of community computer kiosks could greatly improve access to smartphones in rural areas. It is vital that telecommunication regulators enforce the 2% UN Broadband Commission recommendation.

**Funding local language and content programmes:** Funding should be done purposefully on the part of international donors and multinational companies providing the technology, while research funding from within the countries should concentrate more on creating NLP technology tailored for indigenous languages. The government-funded cultural institutions and public broadcasting organisations should create local databases by digitising local content and collaborating with university AI laboratories to create language-specific tools like an education-based chatbot in the Swahili language. Organising AI competitions and hackathons is also one way of encouraging AI development within the community. Improvement can be measured by the number of AI tools created and the amount of African language datasets.

---

[15] https://www.africadatacentres.com/africa-data-centres-unveils-new-10mw-data-centre-in-lagos/
[16] https://konza.go.ke/
[17] https://www.graphic.com.gh/news/education/one-teacher-one-laptop-distribution-completed-ka-technologies.html
[18] https://www.safaricom.et/en/personal/packages/data

**Building robust human capital for systemic educational reforms:** Educational ministries should make it mandatory for students attending universities and vocational training centres to receive instruction on the basics of artificial intelligence and machine learning right from the beginning. Universities and polytechnics need urgent funding to establish specialised graduate programs in the field of artificial intelligence. Multinational institutions and tech companies should sponsor scholarships aimed at grooming AI professionals in Africa, and this scholarship scheme should have an explicit directive to bridge the digital gender gap. In addition, institutions should also encourage the participation of diasporas through teaching or mentoring of local talents. Tracking the number of AI-related degrees and certificates awarded annually, as well as the ratio of AI-trained professionals to the general population, will indicate success in this area.

**Strengthening research and innovation ecosystems:** Governmental and development organisations need to set up specific grant funds only for applied research on AI technologies that can tackle the issues faced at the local level, including diagnostics, climate adaptation, and agricultural technology. Research institutions also need to create innovation centres where businesses can emerge from research conducted at universities. In order to make research applicable commercially faster, universities need to make their procedures for partnerships with industries very simple and easy to comply with, especially with regard to intellectual property laws and the ethics approval process. The metrics to monitor the goal include the number of academic papers and patents related to AI in Africa per year.

**Improving policy and governance:** There is a need for legislators to ensure that developed national AI strategies are turned into properly funded and actionable implementation plans. The governments should set up multi-stakeholder AI council bodies for overseeing the strategic implementation and enforcement of ethics. The regulatory authorities should offer legal clarifications on issues related to data protection, cross-border data transfers, and the fairness of algorithms in order to boost public confidence and foreign investments. It would be useful for policymakers to employ continental collaboration through the African Union and benchmarking resources such as the Oxford AI Readiness Index in formulating continental-level AI policies. Lastly, there should be policies that encourage private investments by using measures such as innovation vouchers and state procurement programs of locally developed AI technologies. Evaluating this progress will depend on the number of countries with updated, actionable AI strategies and the volume of private and foreign direct investment into the AI sector.

## 7. Conclusion

The emergence of artificial intelligence is undoubtedly an inflexion point in global development. Nevertheless, as elaborated throughout this paper, the ability of Africa to take part in this new era is significantly hindered by a "divide" that stretches well past the lack of state-of-the-art machine learning models. In this case, the "divide" refers to structural deficiencies in terms of infrastructure, human capacity, and institutional changes. Given the enormous share of the population without access to stable broadband services and the continent's minuscule share of world data centre capacity (less than 1%), the physical infrastructure needed to foster an AI economy is seriously lacking. Additionally, this problem is compounded by extremely high costs associated with data transfer and use, widespread digital illiteracy, and insufficient amounts of AI models able to operate with native African languages.

But to view Africa through this lens of divides masks an important and dynamic counter-narrative. The continent is home to a uniquely resilient, grassroots technological awakening. The African technology ecosystem is uniquely positioned to build hyper-localised AI solutions, as evidenced by the innovative, community-led initiatives such as the Masakhane NLP consortium and the pragmatic, problem-solving focus of startups such as PBR Life Sciences. These efforts are actively tackling urgent regional issues in the areas of healthcare, agriculture and financial inclusion. It is impossible to deny the unfettered entrepreneurial and intellectual capacity of the continent, but it is currently blighted by a severe brain drain, limited institutional funding and the hyper-concentration of venture capital within a few Tier-1 tech hubs.

Ultimately, resolving the African AI divide will require nothing short of a paradigm shift from the current state of affairs. This will entail concerted efforts from a variety of actors to ensure that access to digital connectivity and computing power is treated as a public utility rather than a luxury good. Indeed, the implementation of the proposed strategies, whereby efforts will have to move from the aggressive development of affordable broadband networks and sovereign data centres to the actual creation of ecosystems around STEM education and locally produced AI technologies, will be necessary. Otherwise, the introduction of AI could become a catalyst for widening the gap between developed and underdeveloped nations, trapping Africa in a continuous cycle of dependence on other nations for its technological advancements. On the contrary, with adequate planning and investment, there exists the chance for Africa to circumvent many of the technological barriers that currently exist. This way, Africa would progress from being a mere consumer of alien digital technology to a producer of its own.

## References

Acquaye, N. A. (2025, May 24). *Bridging Africa's Digital Divide: $100 Billion Needed to Connect 1 Billion People by 2030*. Linkedin. https://www.linkedin.com/pulse/bridging-africas-digital-divide-100-billion-needed-1-appiah-acquaye-iya4e

Agbeyangi, A., & Suleman, H. (2024). Advances and Challenges in Low-Resource-Environment Software Systems: A Survey. *Informatics 2024, Vol. 11, Page 90*, *11*(4), 90. https://doi.org/10.3390/INFORMATICS11040090

Agwor, A., Melo Zurita, M. de L., & Munro, P. G. (2025). Underground urbanism in Africa: Splintered subterranean space in Lagos, Nigeria. *Urban Studies*, *62*(3), 452–468. https://doi.org/10.1177/00420980231174996;CTYPE:STRING:JOURNAL

Ahmed, A. B., King, B. D., Hiran, K. K., Dadhich, M., & Malcalm, E. (2025). Half a Decade of Artificial Intelligence in Education in Africa: Trends, Opportunities, Challenges and Future Directions. *Journal of Engineering Education Transformations*, *38*(3), 81–100. https://doi.org/10.16920/JEET/2024/V38I3/24246

Aryee, J. N. A., Davies, P., Torsah, G. A., & Amekudzi, L. K. (2026). Rethinking AI Accessibility in African Higher Education Institutions Through Inclusive Participation and Policy Reform . In J. M. L. Ferres (Ed.), *Degrees of Change* (pp. 94–112). Wiley.

https://doi.org/10.1002/9781394413096.CH8;JOURNAL:JOURNAL:BOOKS;WGROUP:STRING:PUBLICATION

Ashiru, G. (2025). *Transforming Healthcare Data and AI Driven Growth in Emerging Markets*. Tech In Africa News. https://www.techinafrica.com/transforming-healthcare-data-and-ai-driven-growth-in-emerging-markets/

Ashok, M., Madan, R., Joha, A., & Sivarajah, U. (2022). Ethical framework for Artificial Intelligence and Digital technologies. *International Journal of Information Management*, *62*, 102433. https://doi.org/10.1016/J.IJINFOMGT.2021.102433

Asiedu, M., Haykel, I., Dieng, A., Kauer, K., Ahmed, T., Ofori, F., Chan, C., Pfohl, S., Rostamzadeh, N., & Heller, K. (2025). Nteasee: Understanding Needs in AI for Health in Africa-A Mixed-Methods Study of Expert and General Population Perspectives. *ACMF AccT 2025 - Proceedings of the 2025 ACM Conference on Fairness, Accountability,and Transparency*, 2363–2436. https://doi.org/10.1145/3715275.3732160;CSUBTYPE:STRING:CONFERENCE

Chair, C. (2026, March 24). *The value of inclusion of African Languages in AI*. Masakhane Blog. https://www.masakhane.io/masakhane-african-languages-hub/updates/the-value-of-inclusion-of-african-languages-in-ai

CIPIT. (2025). The state of AI in Africa report. In *The Centre for Intellectual Property and Information Technology Law (CIPIT)*. https://aiconference.cipit.org/documents/the-state-of-ai-in-africa-report.pdf

Collins, O. (2025, August 24). *Today in Numbers: 27.4M Kenyans are Active Internet Users | by Ocheeky Collins | Medium*. Medium. https://medium.com/@collinsogomboochiko/today-in-numbers-27-4m-kenyans-are-active-internet-users-8885fec7b892

CRE. (2022, November 1). *Infographic: Internet Penetration in Africa*. News. https://www.cre.vc/news-slider/internet-penetration-in-africa-infographic

Datacenters. (2026a, April 26). *Africa Data Centers - 259 Facilities from Operators*. https://www.datacentermap.com/africa/

Datacenters. (2026b, April 27). *Western Europe Data Centers - 2836 Facilities*. Data Center Map. https://www.datacentermap.com/western-europe/

Datacenters. (2026c, May 26). *Egypt Data Centers - 13 Facilities from 19 Operators*. DataCenterMap. https://www.datacentermap.com/egypt/

de-Lima-Santos, M. F., Yeung, W. N., & Dodds, T. (2024). Guiding the way: a comprehensive examination of AI guidelines in global media. *AI & SOCIETY 2024 40:4*, *40*(4), 2585–2603. https://doi.org/10.1007/S00146-024-01973-5

Department of Communications and Digital Technologies. (2026). *Draft South Africa National Artificial Intelligence (AI) Policy (Notice 3880 of 2026, Government Gazette No. 54477)* (Number 54477). https://www.gpwonline.co.za

Ehimuan, B., Anyanwu, A., Olorunsogo, T., Akindote, O. J., Abrahams, T. O., & Reis, O. (2024). Digital inclusion initiatives : Bridging the connectivity gap in Africa and the USA – A review. *International Journal of Science and Research Archive*, *11*(1), 488–501.

Falowo, O., & Falowo, S. (2026). Low Internet Penetration in Sub-Saharan Africa and the Role of LEO Satellites in Addressing the Issue. *Telecom 2026, Vol. 7, Page 7*, *7*(1), 7. https://doi.org/10.3390/TELECOM7010007

Feijóo, C., Kwon, Y., Bauer, J. M., Bohlin, E., Howell, B., Jain, R., Potgieter, P., Vu, K., Whalley, J., & Xia, J. (2020). Harnessing artificial intelligence (AI) to increase wellbeing for all: The case for a new technology diplomacy. *Telecommunications Policy*, *44*(6), 101988. https://doi.org/10.1016/J.TELPOL.2020.101988

Filipenco, D. (2026, January 30). *Top 20 countries by the number of data centers in 2025 DevelopmentAid*. Developmentaid . https://www.developmentaid.org/news-stream/post/204016/top-20-countries-by-the-number-of-data-centers-in-2025

Folorunso, A., Olanipekun, K., Adewumi, T., & Samuel, B. (2024). A policy framework on AI usage in developing countries and its impact. *Global Journal of Engineering and Technology Advances*, *21*(1), 154–166. https://doi.org/10.30574/GJETA.2024.21.1.0192

Gala, P. M., Namit, K., & Kidwai, H. (2025). *Strategies for Enhancing Digital Skills among Africa's NEET Youth* (Number 13).

GSMA. (2024). *The mobile economy Sub-Saharan Africa 2024*. https://www.gsma.com/mobileeconomy/wp-content/uploads/2024/05/GSMA_ME_SSA_2024_Web.pdf

ITU. (2023). Measuring Digital Development: Facts and figures. In *ITU Publications*. https://www.itu.int/en/mediacentre/Documents/MediaRelations/ITU Facts and Figures 2019 - Embargoed 5 November 1200 CET.pdf

ITU. (2024a). Measuring digital development: Facts and figures. In *International Telecommunication Union*. International Telecommunication Union. https://www.itu.int/itu-d/reports/statistics/2024/11/10/ff24-internet-use/?utm_source=chatgpt.com

ITU. (2024b, November 10). *Facts and Figures 2024 - Mobile network coverage*. Facts and Figures 2024. https://www.itu.int/itu-d/reports/statistics/2024/11/10/ff24-mobile-network-coverage/

ITU. (2025a). *AI for Good Impact Report* (2nd ed.). International Telecommunication Union, Telecommunication Standardization Sector. https://aiforgood.itu.int/newsroom/publications-and-reports/

ITU. (2025b). *Facts and Figures 2025: Mobile network coverage*. Facts and Figures. https://www.itu.int/itu-d/reports/statistics/2025/10/15/ff25-mobile-network-coverage/

ITU. (2025c). Measuring Digital Development: Facts and figures. In *ITU Publications*. https://www.itu.int/en/mediacentre/Documents/MediaRelations/ITU Facts and Figures 2019 - Embargoed 5 November 1200 CET.pdf

ITU. (2025d). *Over four in five people worldwide own a mobile phone*. https://www.itu.int/itu-d/reports/statistics/2025/10/15/ff25-mobile-phone-ownership/

Jones, B. (2026, May 8). *South Africa launches military-focused AI hub to harness emerging technologies*. Security and Fire Africa News. https://securityafricamagazine.com/south-africa-launches-military-focused-ai-hub-to-harness-emerging-technologies/

Kabengele, C., & Hahn, R. (2025). Venture capital funding in Africa: a mixed-methods study of evolving ecosystems and financial discrimination. *Journal of International Business Studies 2025 56:6*, *56*(6), 777–794. https://doi.org/10.1057/S41267-025-00788-W

Kanyesigye, M. R. (2026). Rethinking Brain Drain in Africa: Factors Driving the Exodus of the Highly Educated and the Role of Higher Education in Reversing the Trend. *Migration and Development*. https://doi.org/10.1177/21632324251352312

Kato, A. I. (2026). Venture capital as Institutional Infrastructure: Modelling the pathways to scalability and competitiveness in high-growth ventures in Uganda. *Future Business Journal 2026 12:1*, *12*(1), 142-. https://doi.org/10.1186/S43093-026-00834-5

Kemp, S. (2025a, November 8). *Digital 2026: Egypt — DataReportal – Global Digital Insights*. Datareportal. https://datareportal.com/reports/digital-2026-egypt?rq=Egypt

Kemp, S. (2025b, November 8). *Digital 2026: Morocco — DataReportal – Global Digital Insights*. Datareportal. https://datareportal.com/reports/digital-2026-morocco?rq=Morocco

Kemp, S. (2025c, November 8). *Digital 2026: Nigeria — DataReportal – Global Digital Insights*. Datareportal. https://datareportal.com/reports/digital-2026-nigeria

Kemp, S. (2025d, November 8). *Digital 2026: South Africa — DataReportal – Global Digital Insights*. Datareportal. https://datareportal.com/reports/digital-2026-south-africa

Khan, M. S., Umer, H., & Faruqe, F. (2024). Artificial intelligence for low income countries. *Humanities and Social Sciences Communications 2024 11:1*, *11*(1), 1422-. https://doi.org/10.1057/s41599-024-03947-w

Lechman, A. (2026, January 5). *Urgent Need for AI Skills Development Accelerates Across Africa - SAP Africa News Center*. SAP News Center. https://news.sap.com/africa/2026/01/urgent-need-for-ai-skills-development-accelerates-across-africa/

Madikane, S. (2026, May 11). *From side hustle to digital powerhouse: Bridging Africa's $100bn AI gap*. Google Africa Blog. https://blog.google/intl/en-africa/from-side-hustle-to-digital-powerhouse-bridging-africas-100bn-ai-gap/

Malatji, M. (2026). *Bridging the AI divide in sub-Saharan Africa: Challenges and opportunities for inclusivity*. https://arxiv.org/pdf/2601.06145

Mannuru, N. R., Shahriar, S., Teel, Z. A., Wang, T., Lund, B. D., Tijani, S., Pohboon, C. O., Agbaji, D., Alhassan, J., Galley, J. Kl., Kousari, R., Ogbadu-Oladapo, L., Saurav, S. K., Srivastava, A., Tummuru, S. P., Uppala, S., & Vaidya, P. (2025). Artificial intelligence in developing countries: The impact of generative artificial intelligence (AI) technologies for development. *Information Development*, *41*(3 Special Issue: "Artificial Intelligence initiatives"), 1036–1054. https://doi.org/10.1177/02666669231200628;PAGE:STRING:ARTICLE/CHAPTER

Manouan, C., Anagbah, M., Douha, N. Y.-R., & Barros, J. (2026). *Take the Train: Africa at the Crossroad of Modern AI*. https://arxiv.org/pdf/2603.21795

Mediacentre. (2024, October 26). *Global Internet use continues to rise but disparities remain, especially in low-income regions*. Press Release. https://www.itu.int/en/mediacentre/Pages/PR-2024-11-27-facts-and-figures.aspx

Mkalama, B., & Ouma, S. (2025). Making sense of funding inequalities in the venture capital space: a state of the art review paper with views from Africa. *Socio-Economic Review*, *23*(2), 1013–1037. https://doi.org/10.1093/SER/MWAF007

Mpedi, L., Maluleke, T., Marwala, T., & Ntshavheni, K. (2022, December 6). *South Africa's new National Artificial Intelligence Institute can help transform our economy*. University of Johannesburg News. https://news.uj.ac.za/news/south-africas-new-national-artificial-intelligence-institute-can-help-transform-our-economy-2/

Mulungu, K. (2025). Gender perspectives on mobile phone ownership and use: a case study of smallholder farmers in Uganda. *Gender, Technology and Development*, *29*(3), 384–397. https://doi.org/10.1080/09718524.2025.2557649

Mwema, E., & Birhane, A. (2024). Undersea cables in Africa: The new frontiers of digital colonialism. *First Monday*, *29*(4). https://doi.org/10.5210/FM.V29I4.13637

Nye, B. D., & Nye, B. D. (2014). Intelligent Tutoring Systems by and for the Developing World: A Review of Trends and Approaches for Educational Technology in a Global Context. *International Journal of Artificial Intelligence in Education 2014 25:2*, *25*(2), 177–203. https://doi.org/10.1007/S40593-014-0028-6

OECD. (2025). *OECD Capital Market Series Africa Capital Markets Report 2025*. https://www.oecd.org/en/publications/2025/11/africa-capital-markets-report-2025_a973e07d/full-report/harnessing-ai-in-finance-for-financial-inclusion-in-africa_a048b4fb.html

OECD. (2026). Venture capital investments in artificial intelligence through 2025. In *Policy Brief*. https://doi.org/10.1007/s00191-024-00857-7

OECD.AI. (2025, December 25). *Launch of the Artificial Intelligence (AI) Institute of South Africa and AI Hubs (University of Johannesburg and Tshwane University of Technology)*. OECD AI Initiatives. https://oecd.ai/en/dashboards/policy-initiatives/launch-of-the-artificial-intelligence-ai-institute-of-south-africa-and-ai-hubs-university-of-johannesburg-and-tshwane-university-of-technology-9180

O'grady Vaughan. (2024, September 16). *Two new initiatives aim to address East African digital inequality*. Developing Telecoms - Investment News. https://developingtelecoms.com/telecom-business/telecom-investment-mergers/17317-two-new-initiatives-aim-to-address-east-african-digital-inequality.html

Okolo, C. T. (2026, April 1). *Strategic Geopolitical Competition and Africa's AI Future*. Just Tech. Social Science Research Council. https://doi.org/10.35650/JT.3095.D.2026

Oleet, J. B. (2025). The Role of Artificial Intelligence in Economic Development. In R. S. Gwala (Ed.), *Driving Socio-Economic Growth With AI and Blockchain* (pp. 339–364). IGI Global Scientific Publishing. https://doi.org/10.4018/979-8-3693-8664-4.CH015

Oloyede, J. (2025a, June 27). *Agentic AI could be Africa's next big competitive advantage*. Techcabal Insights. https://insights.techcabal.com/agentic-ai-could-be-africas-next-big-competitive-advantage/

Oloyede, J. (2025b, July 6). *Africa's AI divide: The three tiers of digital readiness*. Techcabal Insights. https://insights.techcabal.com/africas-ai-divide-the-three-tiers-of-digital-readiness/

Omoniyi, F. (2024, December 20). *Techstars-backed PBR Life Sciences raises $1 million pre-seed funding*. Techcabal News. https://techcabal.com/2024/12/20/pbr-life-sciences-raises-1-million-pre-seed/

O'Neill, J., Marivate, V., Glover, B., Karanu, W., Tadesse, G. A., Gyekye, A., Makena, A., Rosslyn-Smith, W., Grollnek, M., Wayua, C., Baguma, R., Maduke, A., Spencer, S., Kandie, D., Maari, D. N., Mutangana, N., Axmed, M., Kamau, N., Adamu, M., … Nyairo, S. (2024). *AI and the Future of Work in Africa White Paper*. https://arxiv.org/pdf/2411.10091

PBR Life Sciences. (2026, March 7). *Pharmaceutical Market Intelligence in Africa: How Data Analytics Helped a Global Consumer Health Company Evaluate the OTC Vitamins Market*. Case Study. https://pbrinsight.com/insights/

Pouris, A., & Thirion, A. (2024). The State of Artificial Intelligence Research in SADC Region - A Scientometric Assessment. *International Conference on Electrical and Computer Engineering Researches, ICECER 2024*. https://doi.org/10.1109/ICECER62944.2024.10920372

Praxis. (2026). *Praxis Portfolio Ventures | PBR Life Sciences*. Portfolio. https://www.praxis.co/ventures/pbr-life-sciences

Qaleon. (2026, April 27). *AI Investment in 2026: Historic Record and the Rise of AI in Defense*. Blog. https://qaleon.com/en/blog/ai-investment-in-2026-historic-record-and-the-rise-of-ai-in-defense/

Robinson, L., Schulz, J., Blank, G., Ragnedda, M., Ono, H., Hogan, B., Mesch, G., Cotton, S. R., Kretchmer, S. B., Hale, T. M., Drabowicz, T., Yan, P., Wellman, B., Harper, M. G., Quan-Haase, A., Dunn, H. S., Casilli, A. A., Tubaro, P., Carveth, R., … Khilnani, A. (2020). Digital inequalities 2.0: legacy inequalities in the information age. *First Monday*, *25*(7). https://doi.org/10.5210/FM.V25I7.10842

Sajid, I. (2025). *AI for impact at scale: Case studies from innovators in low- and middle-income countries*. GSMA. https://www.gsma.com/solutions-and-impact/connectivity-for-good/mobile-for-development/gsma_resources/ai-for-impact-at-scale-in-lmics/

Segun, S. T., Adams, R., Florido, A., Timcke, S., Shock, J., Junck, L., Adeleke, F., Grossman, N., Alayande, A., Kponyo, J. J., Smith, M., Fosu, D. M., Tetteh, P. D., Arthur, J., Kasaon, S., Ayodele, O., Badolo, L., Plantinga, P.,

Gastrow, M., … Bissyande, T. (2025). *Toward an African Agenda for AI Safety*. https://doi.org/10.1007/s00146-019-00887-x

Selemela, A., Khwela, M. N., Selelo, M. E., Selemela, A., Khwela, M. N., & Selelo, M. E. (2023). The role of the fourth industrial revolution in achieving economic development: challenges and opportunities. *International Journal of Research in Business and Social Science (2147-4478)*, *12*(7), 244–253. https://doi.org/10.20525/IJRBS.V12I7.2808

Shanahan, M. (2025, October 28). *Despite improvements, Sub-Saharan Africa has the widest usage and coverage gaps worldwide | Mobile for Development | GSMA*. GSMA. https://www.gsma.com/solutions-and-impact/connectivity-for-good/mobile-for-development/blog/despite-improvements-sub-saharan-africa-has-the-widest-usage-and-coverage-gaps-worldwide/

Slator. (2026, May 18). *New Initiative Supports Language AI Projects in Africa*. Slator News. https://slator.com/new-initiative-language-ai-projects-africa/

Stewart, G. (2025, May 26). *AI Skills Development in Africa – New Report Findings Revealed - SAP Africa News Center*. SAP News Center. https://news.sap.com/africa/2025/05/ai-skills-development-in-africa-new-report-findings-revealed/

UNESCO. (2024, November 26). *What you need to know about the challenges of STEM in Africa | UNESCO*. News. https://www.unesco.org/en/articles/what-you-need-know-about-challenges-stem-africa

UNESCO. (2025, May 10). *Number of students in higher education more than doubled in 20 years*. Press Release. https://www.unesco.org/en/articles/number-students-higher-education-more-doubled-20-years-inequalities-remain

Unver, M. B. (2025). Democratic AI governance: framing a vision for Africa in view of the EU experience. *International Review of Law, Computers & Technology*. https://doi.org/10.1080/13600869.2025.2506924

Velasco, L., Adan, S. N., Trager, R., Khan, M. S., Fox, J., Corona, R., Adeleke, F., Effoduh, J. O., Eder, L., Sharp, M., Muhaj, D., Kalcic, K., & Trager, R. (2026). *Financing the AI Triad : Compute , Data and Algorithms* (Number February).

Walter, Y. (2024). Managing the race to the moon: Global policy and governance in Artificial Intelligence regulation—A contemporary overview and an analysis of socioeconomic consequences. *Discover Artificial Intelligence 2024 4:1*, *4*(1), 14-. https://doi.org/10.1007/S44163-024-00109-4

Wawire, B. A., Wawire, G. N., & Kiroro, F. (2025). The Structural Relations of Component Reading Comprehension Skills in Kiswahili: The Influence of Socio-Economic Status and Home Literacy Environment. *Reading Research Quarterly*, *60*(1), e593. https://doi.org/10.1002/RRQ.593;WEBSITE:WEBSITE:ILA;WGROUP:STRING:PUBLICATION

Wereko-Brobby, N. (2025, April 29). *How this African startup is using AI to close data gaps in healthcare*. Google - Outreach and Initiatives News. https://blog.google/intl/en-africa/company-news/outreach-and-initiatives/the-kenyan-startup-using-ai-to-close-data-gaps-in-healthcare/

World Bank Group. (2023, June 27). *From Connectivity to Services: Digital Transformation in Africa*. Results Brief. https://www.worldbank.org/en/results/2023/06/27/from-connectivity-to-services-digital-transformation-in-africa

World Bank Group. (2025). *Digital Progress and Trends Report 2025: Strengthening AI Foundations*. International Bank for Reconstruction and Development / The World Bank. https://hdl.handle.net/10986/43822.

Xulu, K. K., & Malatji, M. (2025). The Socioeconomic and Technological Implications of 5G Deployment in Urban and Rural Communities of South Africa. *2nd International Conference on IT Innovations and Knowledge Discovery, ITIKD 2024*. https://doi.org/10.1109/ITIKD63574.2025.11004726